\documentclass[twocolumn,showpacs,prl,amsmath,amssymb,floatfix,superscriptaddress]{revtex4}

\usepackage{graphicx}
\usepackage{dcolumn}
\usepackage{bm}

\begin{document}

\preprint{APS/123-QED}

\title{Exciton condensate at a total filling factor of 1 in Corbino 2D electron bilayers}

\author{L. Tiemann}
\author{J. G. S. Lok}
\author{W. Dietsche}
\author{K. von Klitzing}
\affiliation{ Max-Planck-Institut f\"{u}r Festk\"{o}rperforschung,
Heisenbergstra{\ss}e
1, 70569 Stuttgart, Germany}%

\author{K. Muraki}
\affiliation{
NTT Basic Research Laboratories, NTT Corporation, 3-1 Morinosato-Wakamiya, Atsugi 243-0198, Japan}

\author{D. Schuh}
\author{W. Wegscheider}
\affiliation{ Fakult\"{a}t f\"{u}r
Physik, Universit\"{a}t Regensburg, 93040 Regensburg, Germany}

\date{\today}

\begin{abstract}
Magneto-transport and drag measurements on a quasi-Corbino 2D
electron bilayer at the systems total filling factor 1 ($\nu_T$=1)
reveal a drag voltage that is equal in magnitude to the drive
voltage as soon as the two layers begin to form the expected
$\nu_T$=1 exciton condensate. The identity of both voltages remains
present even at elevated temperatures of 0.25~K. The conductance of
the drive layer vanishes only in the limit of strong coupling
between the two layers and at T$\rightarrow$0~K which suggests the
presence of an excitonic circular current.
\end{abstract}

\maketitle

When two closely spaced two-dimensional electron systems (electron
bilayer) are exposed to a perpendicular magnetic field $B$ so that
each layer has a filling factor close to 1/2 and the relative
distance between interlayer electrons, parameterized by the ratio
$d/l_B$ ($d$: layer separation, $l_B=\sqrt{\hbar /eB}=1/\sqrt{2\pi
n_T}$: magnetic length with $n_T$ as the total density), is
sufficiently small, a new quantum Hall (QH) state characterized by
the total filling factor 1 ($\nu_T$=1) can be observed. Both
theoretically and experimentally it is found that this novel state
in bilayers occurs at a $d/l_B$ ratio of less than $\approx$ 2
\cite{Spielmann00,Kellogg02}. In the limit of comparably small
tunneling, its origin is dominated by Coulomb interactions, where
the electrons in the two layers form a strongly correlated many-body
state to minimize their exchange energy. In this state and in the
low temperature limit, spontaneous interlayer phase coherence
develops, driving all electrons in a quantum mechanical
superposition of the layer eigenstates sharing the same macroscopic
phase $\phi $ \cite{MacDonald01,Wen92}. However, the predicted
Kosterlitz-Thouless type of phase-transition \cite{Moon1995} has not
yet been unequivocally demonstrated in experiment. After a
particle-hole transformation that changes the sign of the
interactions from repulsive to attractive, this state can be
regarded as an excitonic condensate, where each electron is bound to
a \textquotedblleft vacant state\textquotedblright\ in the opposite
layer. Interlayer drag experiments, where a constant current is
passed through one of the layers (\textquotedblleft drive
layer\textquotedblright ) and the induced longitudinal and
transverse voltage drop in the other layer (\textquotedblleft drag
layer\textquotedblright ) is measured \cite{Kellogg02}, have
revealed a Hall drive and drag which approaches a quantized value of
$h/e^{2}$ at $\nu_T=1$ in a temperature-activated fashion. The
quantization of the Hall drag is an indirect indication of a
superfluid mode of excitons \cite{Eisenstein03}, which can be viewed
as either an uniform flow of interlayer excitons or as a counter
flow of electrons in the opposite layers.

In standard Hall bars the occurrence of the ordinary integer QH
effect with the vanishing of the longitudinal resistance and the
quantization of the Hall resistance can be explained in terms of
one-dimensional (dissipationless) edge channels \cite{LB85,LB88}.
However, in the case of the $\nu_T=1$ and its associated superfluid
transport mode, it cannot be ruled out from the Hall bar data that a
dissipationless quasi-particle current at the sample edges is
responsible for the observed effects \cite{Rossi2005}.

In this paper, we report on interlayer drag measurements on a
\emph{quasi}-Corbino electron bilayer with independent contacts to
both layers. An \emph{ideal} Corbino structure allows direct
measurement of the conductivity $\sigma_{xx}$ in contrast to the
common Hall bar geometry where the resistivities are measured. We
observe that at $\nu_T=1$ a voltage develops in the drag layer that
equals in sign and magnitude the voltage across the drive layer. We
find that the identity of the drag and drive voltages is maintained
up to high temperatures or large $d/l_B$ where the single layer
current flow shows nearly no trace of the $\nu_T=1$ QH effect. Due
to the absence of sample edges connecting source and drain contacts
in a ring, the current is driven selectively through the bulk of the
$\nu_T=1$ system. At low temperatures, drive and drag voltage remain
identical and the drive current nearly vanishes. The Corbino
experiments thus open a new venue to explore the bulk property of
the $\nu_T=1$ system.

\begin{figure}[tp]
 \includegraphics[width=0.48\textwidth]{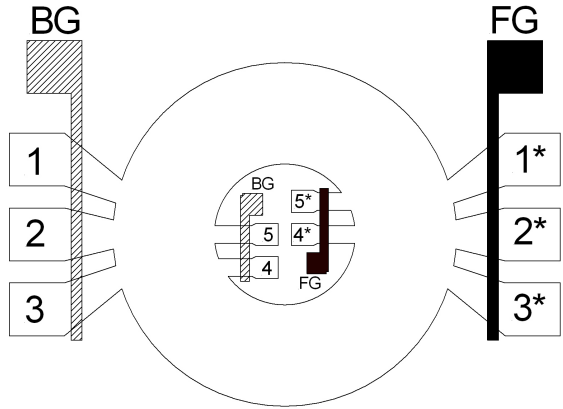}
 \caption{Schematic view of the Corbino geometry used in this experiment.
  Application of appropriate voltages to the back gates (marked as ''BG'')
  and front gates (''FG'') will lead to contact separation, i.e. contacts
  1 through 5 will connect to the upper quantum well and 1* through 5* to
  the lower one.}
 \label{fig:Corbino}
\end{figure}

Our two-dimensional (2D) electron bilayer is confined in two 19-nm
GaAs quantum wells, separated by a 9.9 nm superlattice barrier
composed of alternating layers of AlAs (1.70 nm) and GaAs (0.28 nm).
Each quantum well has an intrinsic electron density of about
$4.3\times 10^{14}$ m$^{-2}$ and a low-temperature mobility of 67
(45) m$^{2}$/Vs for the upper (lower) quantum well (measured on a
Hall bar fabricated from the same wafer). Since the ideal Corbino
geometry is not compatible with the selective-depletion technique
\cite{Eisenstein90, Rubel97} for independently contacting each
layer, we instead employ a quasi-Corbino geometry with four contact
arms attached to each ring as depicted in Fig.~\ref{fig:Corbino}.
The back gates were patterned \emph{ex situ} from a Si-doped GaAs
epilayer before growing an insulating GaAs/AlGaAs superlattice and
the bilayer on top. Electrical isolation between the two layers is
achieved by applying appropriate negative voltages to the buried
back gates and metallic front gates crossing the contact arms. One
set of contacts can then be used to pass a current and another one
to measure the voltage across the ring. The densities in each layer
can be adjusted independently by using another set of front and back
gates (not shown) covering the active region of the Corbino ring
including the ring edges.

Below we present data from two samples from the same wafer which
show essentially the same behavior. Sample A consists of a
quasi-Corbino ring with an outer diameter of $d_O=600$ $\mu $m and a
ring width of $w=140$ $\mu $m, while sample B is characterized by
$d_{O}=780$ $\mu $m and $w=230$ $\mu $m. For all samples interlayer
tunneling is small; the interlayer resistance (at zero magnetic
field and 0.25~K) is of the order $10^{7}\ $-$10^{8}$\thinspace
$\Omega$. Transport measurements were performed by using a standard
lock-in technique with the sample mounted at the cold finger of a
dilution refrigerator or a $^3$He system. For all measurements the
electron densities in the two layers were adjusted to be equal. A
small excitation voltage $V_{exc}$ (60-65\thinspace $\mu $V,
3-5\thinspace Hz \footnote{The excitation voltage is generated with
the internal oscillator of one lock-in amplifier and sized down with
a voltage divider.}) was applied radially across one layer (the
drive layer) through an isolation transformer and the induced
current through this layer was measured with a small resistance
connected in series. We would like to stress that the total current
has a radial and an azimuthal part. These two parts oscillate
anti-cyclically as a function of the magnetic field, i.e. in a QH
state the radial fraction is zero while the azimuthal (circular)
part is maximal. Hence, the (radial) voltage dropping over the drive
layer changes in response to the radial current as well. For that
reason, the voltage across the drive layer was monitored using a
separate pair of contacts in a quasi four-terminal geometry together
with the induced voltage in the drag layer. This excludes also the
effects of the finite resistances of the ohmic contacts and the
contact arms. The measurements were reproducible upon interchanging
contacts and upon interchanging drive and drag layer.

\begin{figure}[tp]
 \includegraphics[width=0.5\textwidth]{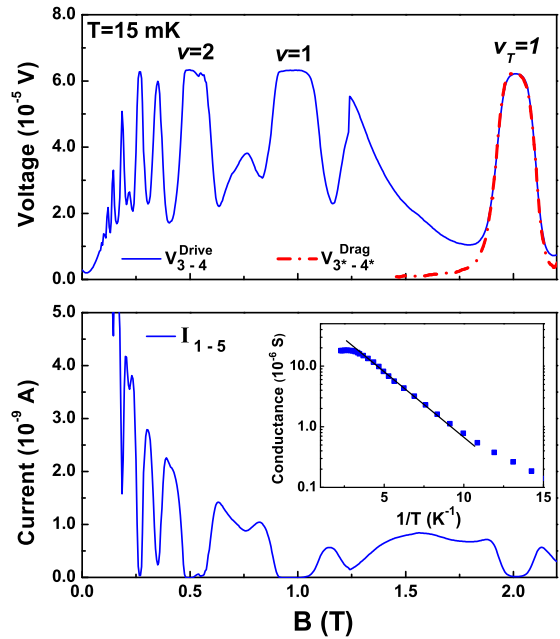}
 \caption{Top panel: Measured drive (solid line) and drag (dash-dot line)
  voltages at $T_{bath}$=15 mK~on sample B. The (integer) filling factors
  $\nu\leq2$ and $\nu_T=1$ ($d/l_B$=1.62) are labeled. Bottom panel: Measured
  current in the drive layer. The inset plots the temperature dependence of
  the radial conductance G; the line is a fit using G~$\propto$~exp(-$E_{gap}$/T).}
 \label{fig:fullsweep1}
\end{figure}

We start by showing data at lowest temperatures.
Fig~\ref{fig:fullsweep1} presents data at $T_{\mathrm{bath}}$ =
15~mK on sample B. The bottom and top panel plot the measured
(radial) current in the drive layer and the corresponding drive and
drag voltages as a function of the magnetic field. The electron
densities in the two layers were adjusted to be equal, producing a
total density of $4.8\times 10^{14}$\thinspace m$^{-2}$. Below
1.5~T, the current oscillates reflecting the varying filling factors
and integer QH states. At total filling factor 1 which occurs here
at about 2.0~T, we observe a strong minimum in the current like at
the ordinary QH states at lower magnetic fields. As a result, the
voltage drop over the drive layer almost equals the source voltage
(top panel). Meanwhile, a large drag voltage develops over the
region of the correlated $\nu_T=1$ phase, with the sign and
magnitude identical to that of the drive layer. Since the radial
component of the current in the drive layer is nearly zero, one
possible explanation for the observed drag voltage is the existence
of an azimuthal (i.e. circling) current in the drive layer, in
analogy to the ordinary QH states. Owing to the excitonic coupling
it would trigger an azimuthal current of the same magnitude in the
drag layer, leading to identical voltages across both layers.
However, we neither know the nature of this excitonic current nor
where it flows. It could be homogeneously distributed throughout the
bulk or rather concentrated at the sample edges. Nevertheless, the
well-established model of electron-hole pairing around $\nu_T=1$
implies that such a transport mode in Corbino bilayers might be
possible. Supported is that notion by the fact that the ohmic
contacts of the drag layer in our geometry are located at the
opposite side of the ring, i.e. approximately 1 mm away from the
ohmic contacts of the drive layer. In previous drag experiments
using a standard Hall bar geometry, identical Hall voltages in the
drag and drive layers were also considered to be signaling the
underlying excitonic superfluidity.

The origin of identical voltages could equivalently be attributed to
the special nature of the excitonic state. Since an excitonic
(quasi-particle) wave function would have to exist across the
barrier, quasi-particle transfer between the layers would become
possible as soon as the system reaches a total filling factor of 1.
While standard tunneling spectroscopy experiments \cite{Spielmann00}
performed on very similar electron bilayer samples indeed indicate
that tunneling becomes resonantly enhanced in the vicinity of
$\nu_T=1$, identical voltages could only be explained if the
interlayer resistance became insignificantly small compared to the
bulk resistance. This, however, is inconsistent with tunneling
experiments \cite{Spielmann00,Spielmann04, Wiersma06} on common
electron bilayer samples, showing resistances within the M$\Omega$
range instead.

\begin{figure}[tp]
 \includegraphics[width=0.5\textwidth]{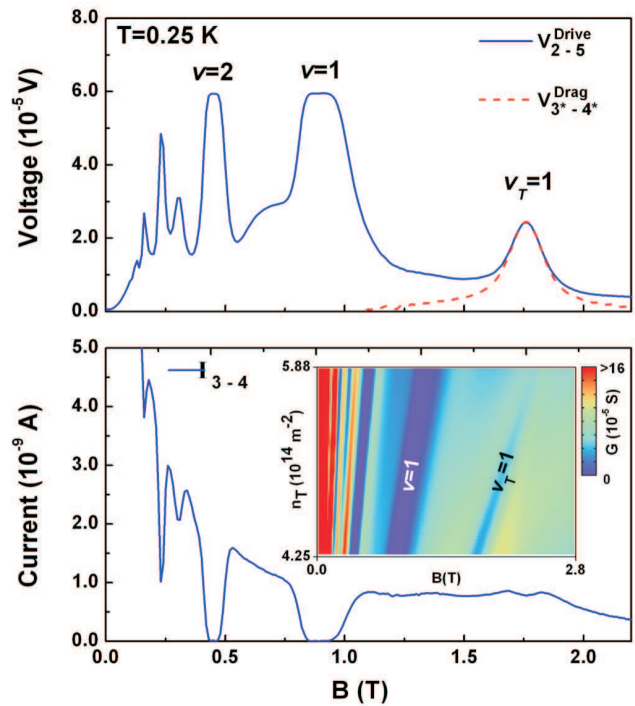}
 \caption{Top panel: Drive (solid line) and drag (dash-dot line) voltage
 versus the magnetic field at T = 0.25~K measured on sample A. Bottom
 panel: The current in the drive layer measured simultaneously. The
 inset illustrates the conductance G of the drive layer as a function
 of the magnetic field and the total density. Clearly visible is how
 the conductance at $\nu_T=1$ decreases as the total density $n_T$ is reduced.}
 \label{fig:fullsweep2}
\end{figure}

Fig.~\ref{fig:fullsweep2} plots data taken at a temperature of
$T$=0.25~K on sample~A. The densities in both layers are still equal
but reduced to a total electron density of approximately $4.2\times
10^{14}$ m$^{-2}$. Now $\nu_T=1$ occurs at $B$=1.76~T which
corresponds to $d/l_B=1.49$. At 0.25~K the minimum in the current
has almost entirely disappeared (bottom panel). Nonetheless, there
is still a sizeable peak in the drive voltage at $\nu_T=1$ (solid
line in the top panel). Surprisingly, the voltage over the drag
layer (dash-dot line) also displays a peak {\em with the same
amplitude}. This striking observation of a nearly doubled
dissipation in the drive layer accompanied by an identical drag
voltage can be interpreted as evidence that both layers are in a
state of commencing interlayer correlation. A previous report on
drag experiments on Hall bar bilayers \cite{Kellogg02} has shown
that identical voltages, i.e. the quantization of the Hall
resistance to $h/e^{2}$, are only observable at lowest temperatures
and low $d/l_B$ ratios when the $\nu_T=1$ QH state is fully
developed. While this is in direct contrast to our data and might
indicate a geometry-dependence, the resilience of the $\nu_T=1$ QH
state to increasing temperatures yet is a behavior reminiscent of
results obtained on bilayer 2-dimensional \emph{hole} gas samples
\cite{Tutuc06} in counter-flow configuration. We cannot offer any
explanation for these similarities, however, it might simply be
owing to the reported interlayer leakage or the larger effective
mass of the holes.

\begin{figure}[tp]
 \includegraphics[width=0.5\textwidth]{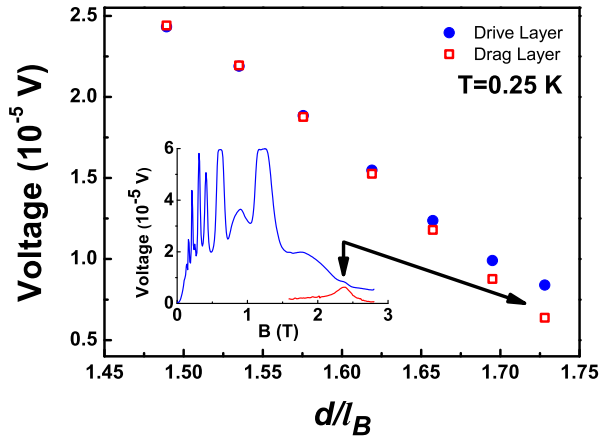}
 \caption{Drive (solid dots) and drag (open dots) voltages at $\nu_T=1$ versus
 the parameter $d/l_B$ at $T_{bath}$ = 0.25 K. Above $d/l_B$=1.65, the amplitudes
 of drive and drag voltage at total filling factor 1 diverge. The inset shows
 the corresponding field sweep for the last pair of points at $d/l_B$=1.73.}
 \label{fig:Vvsdlb}
\end{figure}

We find that the ratio of both voltages remains 1 until $d/l_B$
approaches a critical limit. We have traced the drag and drive
voltages for a number of different (but matched) total densities at
0.25~K. The results are summarized in Fig.~\ref{fig:Vvsdlb} which
plots drag and drive voltage at $\nu_T$=1 versus $d/l_B$. At 0.25~K,
the identity of both drag and drive voltages can be tracked up to a
$d/l_B$ ratio of about 1.65 where the $\nu_T=1$ QH state is
collapsing owing to thermal fluctuations. For $d/l_B>1.65$ small
peaks of different amplitude can be observed as illustrated in the
inset.

At some finite temperature, the collapse of the excitonic condensate
at $\nu_T=1$ in the bilayer can be observed. For sample B and
temperatures below 0.25 K, the conductance G=I/V is well described
by thermal activation, i.e. G~$\propto$~exp(-$E_{gap}$/T), with an
activation energy gap of approximately 0.5 K as shown in the inset
of Fig~\ref{fig:fullsweep1}. The magnitude of the extracted energy
gap is in good agreement with earlier reports on comparable double
quantum well structures
\cite{Kellogg02,Kellogg04,Tutuc04,Wiersma04}, where the activation
energy was extracted from measurements of the temperature dependence
of the longitudinal resistance in Hall bars.

In a theoretical Letter, Stern and Halperin \cite{Stern02} suggested
that the electron bilayer system at high $d/l_B$ ratios is composed
of puddles of strong interlayer correlation incorporated in the
compressible fluids of the individual layers. Their model, albeit
addressing specifically Hall bar geometries, appears to be connected
with our observations as well. As long as these puddles are small in
number and/or unrelated, a sizeable current could flow through the
bulk between source and drain contacts. As $d/l_B$ is decreased
their number and/or size will increase until they eventually
percolate, while the current through the bulk slowly diminishes. The
smooth transition we observe in Corbino samples from a compressible
to a nearly fully incompressible state upon decreasing the
temperature and/or the parameter $d/l_B$ appears to signify such a
percolation.

In conclusion, we have conducted interlayer drag experiments on
quasi-Corbino electron bilayers. At the lowest temperature and
strong coupling, the ratio of drag and drive voltages is 1 while the
conductance in the drive layer vanishes. These data imply a circular
potential distribution along the sample edges owing to a circling
(azimuthal) excitonic current in both layers. At elevated
temperatures, the identity of both voltages is still present.

We thank Allan H. MacDonald for inspiring conversations. Also we
would like to acknowledge Maik Hauser for providing some of our MBE
wafers, J. H. Smet for giving us access to his dilution system, S.
Schmult for his help with the manuscript and the German Ministry of
Education and Research (BMBF) for its financial support (BMBF
01BM456).

\end{document}